\documentclass[lettersize,journal]{IEEEtran}
\usepackage{amsmath,amsfonts}
\usepackage{algorithmic}
\usepackage{algorithm}
\usepackage{array}
\usepackage[caption=false,font=normalsize,labelfont=sf,textfont=sf]{subfig}
\usepackage{textcomp}
\usepackage{stfloats}
\usepackage{float}
\usepackage{url}
\usepackage{verbatim}
\usepackage{graphicx}
\usepackage{cite}
\usepackage{siunitx}
\usepackage{balance}
\usepackage{circuitikz}
\usepackage{capt-of}
\usepackage[export]{adjustbox}
\usepackage{subfig}
\usepackage{hyperref}
\usepackage{color,soul}
\usepackage{titlesec}
\usepackage{orcidlink}
\usepackage{cuted} 
\usepackage{multicol}
\usepackage{mathdots}

\newcommand{\ignore}[1]{}

\makeatletter
\renewcommand*\env@matrix[1][*\c@MaxMatrixCols c]{%
	\hskip -\arraycolsep
	\let\@ifnextchar\new@ifnextchar
	\array{#1}}
\makeatother

\begin{document}
\bstctlcite{IEEEexample:BSTcontrol}

	\title{MEMS Switch Enabled Spatiotemporally Modulated Isolators}
	

        \author{Connor Devitt \orcidlink{0000-0002-3394-6842} ,~\IEEEmembership{Graduate Student Member,~IEEE}, Yong-bok Lee \orcidlink{0000-0002-9539-8298},~\IEEEmembership{Member,~IEEE},\\  Pavitra Jain \orcidlink{0009-0007-0520-2006} ,~\IEEEmembership{Graduate Student Member,~IEEE} Sunil A. Bhave \orcidlink{0000-0001-7193-2241},~\IEEEmembership{Senior Member,~IEEE}, \\ Xu Zhu,~\IEEEmembership{Senior Member,~IEEE}, Nicholas Yost,~\IEEEmembership{Member,~IEEE}, Yabei Gu,~\IEEEmembership{Member,~IEEE}
        \thanks{Manuscript received on XX June 2025; revised on XX XX 2025, accepted on XX XX, 2025. (\textit{Corresponding authors: Connor Devitt, Sunil A. Bhave})}	
        \thanks{Connor Devitt (e-mail: devitt@purdue.edu),  Pavitra Jain (e-mail: jain775@purdue.edu), and Sunil A. Bhave (e-mail: bhave@purdue.edu) are with the OxideMEMS Lab, Elmore Family School of Electrical and Computer Engineering, Purdue University, West Lafayette, IN 47907 USA.}
        \thanks{Yong-bok Lee (e-mail: yblee67@chonnam.ac.kr) is with the Department of Electronics and Computer Engineering at Chonnam National University, Gwangju, South Korea}
        \thanks{Xu Zhu (e-mail: xu.zhu@menlomicro.com) and Nicholas Yost (e-mail: nicholas.yost@menlomicro.com) are with Menlo Microsystems, Inc., Albany, NY 12203 USA.}
         \thanks{Yabei Gu (e-mail: yabei.gu@sandisk.com) is with Sandisk Inc., Milpitas, CA 95035 USA.}}
        
	{}
	
	\maketitle
	

	
	%

\begin{abstract}
This work reports the simulation, design, and implementation of a compact MEMS switch based spatiotemporally modulated (STM) bandpass filtering isolator to improve self-interference cancellation (SIC) in underwater acoustic communication networks. Conventional ferrite circulators are unavailable in ultrasonic frequency ranges limiting SIC to techniques such as spatial cancellation and adaptive digital cancellation. This study details a sub-megahertz electronic non-magnetic filtering isolator. High power-handling, compact, and reliable MEMS switches enable the periodically time varying filter circuit to be non-reciprocal. A printed circuit board (PCB) implementation shows strong agreement with spectral admittance matrix simulations with a maximum measured isolation of 15.99 dB. In conjunction with digital SIC methods, this isolator can enable in-band full duplex underwater communication, environmental sensing, and imaging. 

\end{abstract}

\begin{IEEEkeywords}
Spatiotemporal modulation (STM), MEMS Switches, non-reciprocity, bandpass filter (BPF), underwater communication
\end{IEEEkeywords}

 
\section{Introduction}
 Non-reciprocal radio-frequency (RF) components such as circulators and isolators are essential for a wide range of systems. Circulators are frequently used in RF front-end modules for both half-duplex and full-duplex communication systems providing isolation between the transmit (Tx) and receive (Rx) signal paths \cite{sabharwal_band_2014, reiskarimian_nonreciprocal_2019, choi_achieving_2010, nagulu_nonreciprocal_2018}. They are also critical components in quantum non-demolition superconducting qubit readout \cite{arute_quantum_2019, dassonneville_fast_2020, abdo_high-fidelity_2021, ruffino_wideband_2020, zhuang_superconducting_2024}. Conventionally, reciprocity is broken in these circulators by using ferrite cores with a strong magnetic bias \cite{pozar_microwave_2012, liao1980microwave} resulting in bulky and expensive devices. In quantum applications, the limited space inside dilution refrigerators and the sensitivity to stray magnetic fields further exacerbate the packaging challenges associated with ferrite circulators. Alternatively, reciprocity can be broken without magnetic materials using spatiotemporally modulated (STM) circuits \cite{dutta_spatiotemporal_2022, reiskarimian_nonreciprocal_2019} resulting in significant miniaturization. Non-reciprocal STM filters have been explored using varactor diodes to modulate the capacitance of distributed resonator filters \cite{wu_non-reciprocal_2019, wu_non-reciprocal_2019-1, alvarez-melcon_coupling_2019}, lumped-element filters \cite{wu_isolating_2019, zhang_incorporating_2023, kord_magnet-less_2018, estep_magnetless_2016}, and acoustic resonators \cite{torunbalci_fbar_2018}. However, each varactor requires an additional filter circuit to prevent the signal from leaking out of the modulation path which introduces additional overhead and complexity. Solid state RF switches can provide the modulation with reduced complexity by periodically loading resonators with a capacitance \cite{yu_magnetic-free_2018, pirro_low_2019, khater_switch-based_2024} or switching between acoustic delay lines \cite{lu_radio_2019} and acoustic filters \cite{yu_radio_2019}.

While these STM techniques have been developed primarily for RF ($>\SI{100}{\mega\hertz}$) and superconducting quantum ($4-8$ $\si{\giga\hertz}$) applications as direct replacements for ferrite circulators, there is increasing interest in extending them for use at lower frequencies for applications such as Magnetic Resonance Imaging (MRI) ($1-300$  $\si{\mega\hertz}$) \cite{analog_MRI}, medical ultrasonic imaging ($2-40$  $\si{\mega\hertz}$) \cite{neumann_ultrasound_2018}, and underwater acoustic communication (UWAC) systems ($<\SI{1}{\mega\hertz}$) \cite{cho_survey_2022, stojanovic_recent_1996}, where conventional ferrite circulators are unavailable. UWAC channels are of particular interest and are especially challenging. The water presents a strongly fading channel with frequency-dependent attenuation and exhibits multipath effects, Doppler shifts, long delay spread, time-varying characteristics, and non-Gaussian noise characteristics \cite{stojanovic_underwater_2009}. As a result, carrier frequencies are typically low, often tens to hundreds of kilohertz, communication bandwidths are narrow to accommodate the fading channel and limited transducer bandwidth, and Tx powers are high (often tens of watts), while the received signals are in the milliwatt range. These challenging channel conditions combined with slow propagation speed and disparity between Tx and Rx powers have constrained UWAC systems to operate in non-overlapping frequency and time slots with low bit rates. To improve bit rates and communication efficiency there is significant interest in developing in-band full duplex (IBFD) UWAC systems \cite{hsieh_full-duplex_2023, towliat_self-interference_2020, qiao_research_2021, wang_acoustic-domain_2019, guo_lake_2023, wang_full-duplex_2018}. for a wide range of applications including monitoring marine life, underwater environmental monitoring, and marine exploration.

\begin{figure}[t]
        \centering
        \subfloat[]{\includegraphics[width=0.34\textwidth]{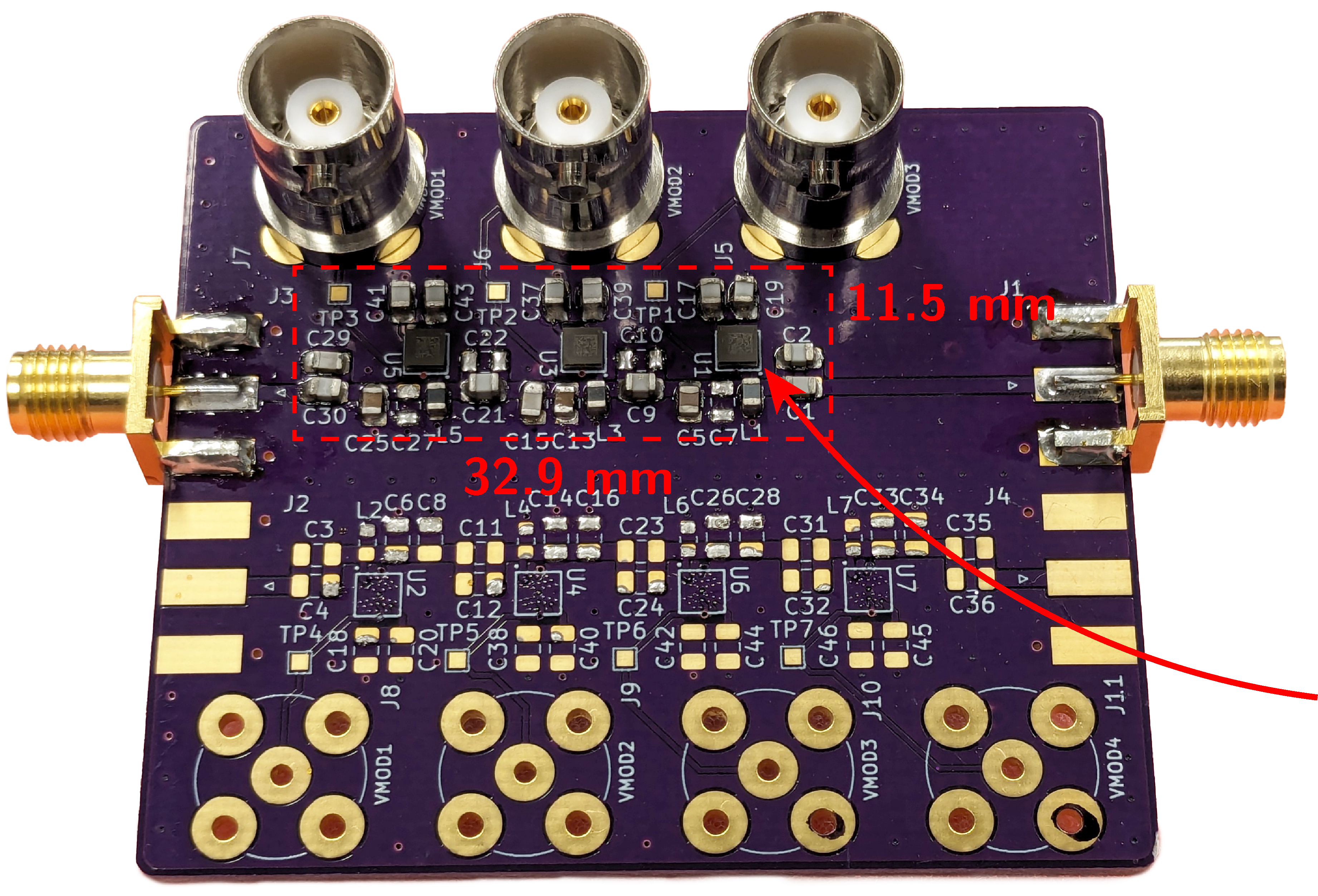}%
        \label{menlo_chip}\hspace*{-0.10in}}%
        \subfloat[]{\includegraphics[width=0.18\textwidth]{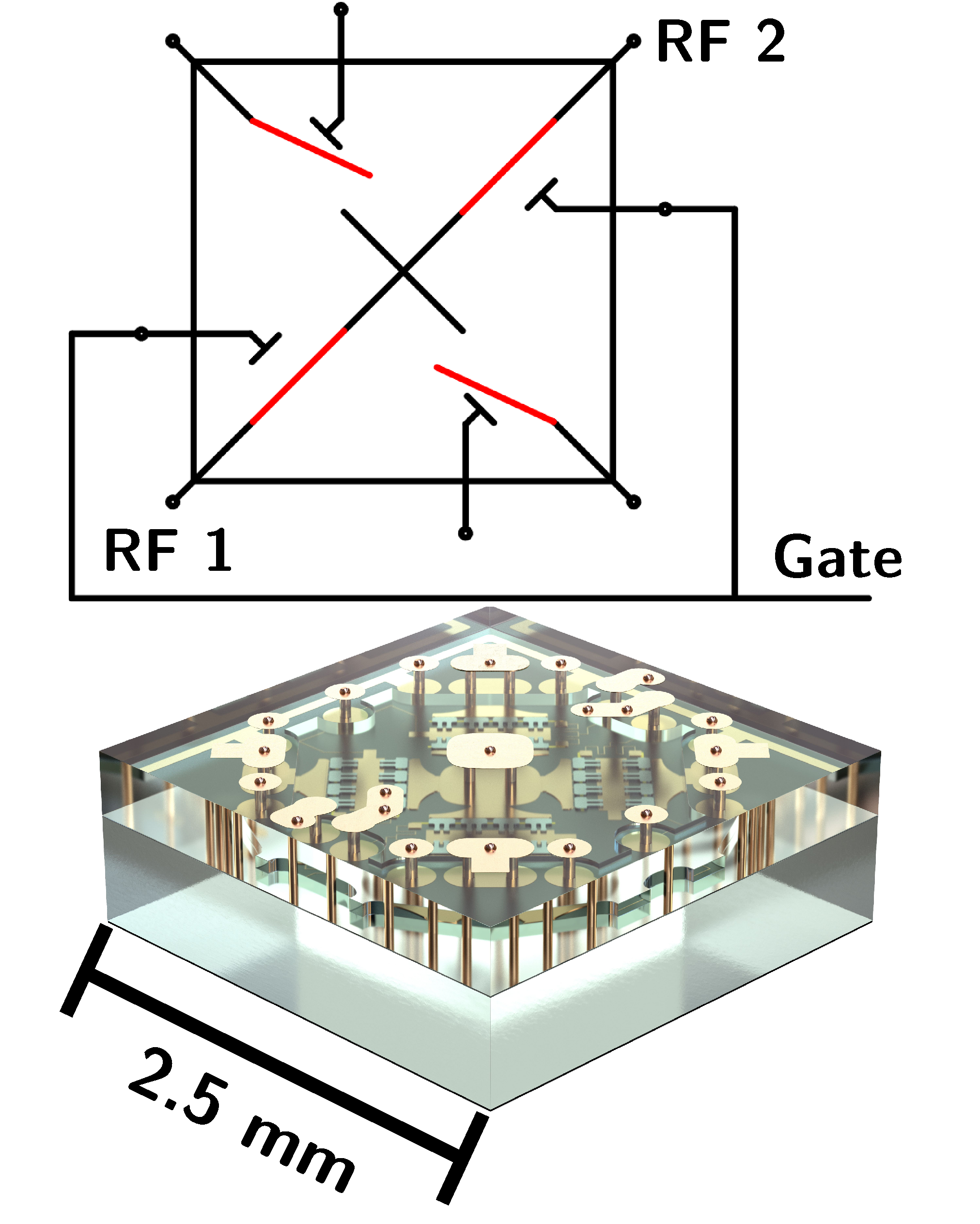}%
        \label{isolator_pcb}}
        \caption{ \textbf{(a)} PCB implementation of a $3^{rd}$-order lumped element STM bandpass isolator using packaged \textbf{(b)} Menlo Microsystems, Inc. MEMS switches. The active isolator area highlighted in red measured $\SI{32.9}{\milli\meter}\times \SI{11.5}{\milli\meter}$. The single-pole, four-throw MEMS switch is configured with two gates shorted together, routing the signal from RF 1 to RF 2 as indicated in \textbf{(b)}. The remaining two throws are unused.}
        \label{device_photo}
    \end{figure}

The primary barrier to IBFD operation, similar to wireless RF applications \cite{sabharwal_-band_2014}, is self-interference (SI) cancellation. Due to the strong multipath effects and high Tx power, SI is significant for UWAC channels. Various SI cancellation methods have been investigated including physical separation of the Tx and Rx transducers, beamforming techniques to suppress SI in the spatial domain \cite{guo_lake_2023,hsieh_full-duplex_2023, wang_acoustic-domain_2019}, and adaptive digital cancellation techniques \cite{qiao_research_2021, wang_full-duplex_2018}. Here, STM non-reciprocal components with high power handling show potential to enable IBFD communication for systems with closely spaced or shared Tx and Rx transducers by improving SI cancellation when combined with previously reported digital SI cancellation techniques. At the carrier frequencies used in UWAC frequencies, microelectromechanical system (MEMS) switches are ideal compact and reliable modulation elements with high power handling capability \cite{micro_rf_nodate}. Moreover, MEMS switches provide excellent isolation ensuring minimal leakage of the signal out of the modulation paths and eliminating the need for any additional filter circuitry. This work demonstrates a spatiotemporally modulated isolator at UWAC frequencies using MEMS switches as pictured in Fig. $\ref{device_photo}$ for self-interference cancellation in underwater channels, thereby enabling in-band full duplex communication.  
    
\section{Isolator Design}

  \begin{figure}[b]
        \centering
            \begin{circuitikz}[scale=0.6]
                \ctikzset{label/align=rotate}
                
                \ctikzset{capacitors/scale=0.4}
                \ctikzset{resistors/scale=0.5}
                \ctikzset{inductors/scale=0.5}
                \ctikzset{bipole label style/.style={font=\footnotesize}}
                
            
                \draw (3,-6) to (3,-6) node[ground]{}; 
                \draw (7,-6) to (7,-6) node[ground]{}; 
                \draw (12,-6) to (12,-6) node[ground]{}; 
            
                \draw (4,-3.5) to[closing switch, switch start arrow=latexslim,
                l_=$t_0$] (4,-2.5);
                \draw (8,-3.5) to[closing switch, switch start arrow=latexslim,
                l_=$t_0  + \Delta t$] (8,-2.5);
                \draw (12,-3.5) to[closing switch, switch start arrow=latexslim,
                l_=$t_0 + 2\Delta t$] (12,-2.5);
                
                \draw
                
                (0,-1) to[C=$C_{k1}$] (3,-1)
                (3,-1) to[C=$C_{k2}$] (7,-1)
                (7,-1) to[C=$C_{k3}$] (11,-1)
                (11,-1) to[C=$C_{k4}$] (14,-1)
                
                (1.75,-6) to[L=$L_{r}$] (1.75,-2)
                (3,-6) to[C=$C_{r1}'$] (3,-2)
                (4,-2) to[short, -o] (4,-2.5)
                (4,-4) to[short, -o] (4,-3.5)
                (4,-6) to[C=$C_{m}$] (4,-4)
                
                (5.75,-6) to[L=$L_{r}$] (5.75,-2)
                (7,-6) to[C=$C_{r2}'$] (7,-2)
                (8,-2) to[short, -o] (8,-2.5)
                (8,-4) to[short, -o] (8,-3.5)
                (8,-6) to[C=$C_{m}$] (8,-4)
                
                (9.75,-6) to[L=$L_{r}$] (9.75,-2)
                (11,-6) to[C=$C_{r3}'$] (11,-2)
                (12,-2) to[short, -o] (12,-2.5)
                (12,-4) to[short, -o] (12,-3.5)
                (12,-6) to[C=$C_{m}$] (12,-4)

                (1.75,-2) -- (4,-2)
                (1.75,-6) -- (4,-6)
                (3,-1) -- (3,-2)
                
                (5.75,-2) -- (8,-2)
                (5.75,-6) -- (8,-6)
                (7,-1) -- (7,-2)
                
                (9.75,-2) -- (12,-2)
                (9.75,-6) -- (12,-6)
                (11,-1) -- (11,-2)
                
                ;
            \end{circuitikz}
        \caption{Circuit diagram for a lumped-element $3^{rd}$-order STM isolator. $L_r$ and $C_r$ form identical resonators and $C_{ki}$ are coupling capacitors chosen to realize a $\SI{0.05}{\decibel}$ Chebyshev filter response. $C_m$ represents the modulation capacitors which are periodically connected to the LC-resonator using Menlo Microsystems switches.}
        \label{filter_schematic}
    \end{figure}
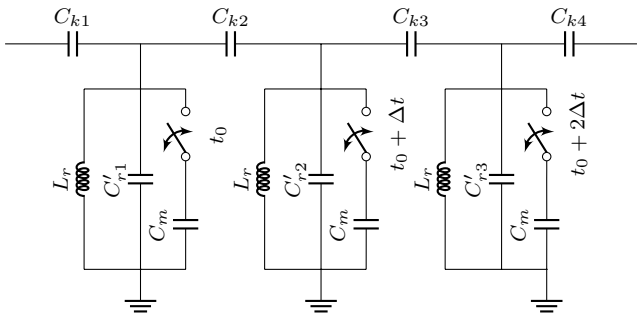
    
Fig. $\ref{filter_schematic}$ shows the circuit schematic for a lossless $3^{rd}$-order STM isolating filter. The isolator is composed of 3 identical lumped element resonators, each of which consists of $L_r$, $C_r$, and an additional modulated capacitance, $C_m$ in series with a MEMS switch. To leverage the design techniques of coupled resonator filters, the total resonator capacitance is modeled as a time-varying capacitor modulated around an effective static capacitance, which for square wave modulation is $\left(C_r + D C_m\right)$ where $D$ is the duty cycle. The time-averaged filter center frequency, $\omega_0$, becomes:
\begin{align}
    \omega_0 &= \frac{1}{\sqrt{L_r \left(C_r + D C_m\right)}}.
    \label{center_freq_eqn}
\end{align}
Each resonator is connected in series with admittance inverters who values, $J_{i}$, are chosen to implement a $\SI{0.05}{\decibel}$ Chebyshev filter with a fractional bandwidth of $\Delta=0.15$ \cite{pozar_microwave_2012}: 
\begin{align}
    J_1 &= \sqrt{\frac{C_r \Delta \omega_0}{Z_0 g_1}}\\
    J_{2\leq i \leq N} &= \Delta\omega_0 \sqrt{\frac{C_r^2}{g_{i-1}g_{i}}}\\
    J_{N+1} &= \sqrt{\frac{C_r \Delta \omega_0}{Z_0 g_{N}g_{N+1}}},
\end{align}
where where $Z_0=50\Omega$ is the characteristic impedance, $N$ is the filter order, and $g_i$ are the filter prototype values for the Chebyshev response.
The interior inverters ($C_{k2}, C_{k3}$) are implemented as a capacitive PI networks with capacitance $C_{ki}=\frac{J_{i}}{\omega_0}$ in the series branch and $-C_{ki}$ in the shunt branches \cite{matthaei1963design, pozar_microwave_2012, wu_isolating_2019, ou_lumped-element_2011}. The first and last inverters ($C_{k1}, C_{k4}$) are implemented as capacitive $L$-networks with series capacitance
\begin{align}
    C_{ki, s} &= \frac{1}{Z_0\omega_0\sqrt{\left(\frac{1}{Z_0 J_i}\right)^2 - 1}},
\end{align}
and shunt capacitance 
\begin{align}
    C_{ki, p} &= -\frac{C_{ki, s}}{1 + \left(\omega_0 C_{ki, s}Z_0\right)^2}.
\end{align}
The negative capacitance values are absorbed by $C_r$ giving the final resonator capacitance $C_{r1}',C_{r2}',$ and $C_{r3}'$ in Fig. $\ref{filter_schematic}$.

The MEMS switches periodically closes according to a square wave with a frequency of $\omega_m$ and duty cycle $D$ which periodically shifts the resonators' frequencies by adding and removing $C_m$ from the circuit. Switches in adjacent resonators close after an incremental time delay $\Delta t$ such that the time varying resonators have a modulation phase progression of $\phi_m$. The value of the modulated capacitor ($C_m$) in series with the switch is chosen through optimization of the modulation strength, $\xi = \frac{C_m}{C_r}$, for the strongest isolation and highest forward transmission.

\begin{figure*}[!t]
    \centering
    \begin{align}
        \label{spectral_admittance}
		Y_{SAM} &= 
			\begin{bmatrix}
				\ddots & \vdots & \vdots & \vdots & \iddots\\
				\cdots &\left(\omega-\omega_m\right)&a_{-1}\xi e^{j\phi_m}\left(\omega-\omega_m\right)&a_{-2} \xi e^{j2\phi_m}\left(\omega-\omega_m\right)& \cdots \\
				\cdots &a_{1} \xi e^{-j\phi_m}\omega&\omega& a_{-1}\xi e^{j\phi_m}\omega& \cdots \\
				\cdots &a_{2} \xi e^{-j2\phi_m}\left(\omega+\omega_m\right)&a_{1} \xi e^{-j\phi_m}\left(\omega+\omega_m\right)&\left(\omega+\omega_m\right)& \cdots\\
				\iddots  & \vdots & \vdots & \vdots & \ddots \\
			\end{bmatrix}
	\end{align}
    \vspace{-0.1in}
\end{figure*}

The spatiotemporally modulated lumped element isolator is simulated either using a harmonic balance simulation in a circuit simulator such as Keysight ADS or using a spectral admittance matrix (SAM)\cite{kurth_steady-state_1977, wu_isolating_2019} method. Both methods produce identical isolator responses for a modulation duty cycle of $50\%$. In the harmonic balance simulation, the Menlo switch in series with $C_m$ is represented by the MEMS switch's on-resistance, $R_{switch}$, in series with voltage-dependent capacitor controlled with a $50\%$ duty-cycle square wave at $\omega_m$with the capacitance periodically alternating between $0$ and $C_m$. 

While the harmonic balance is simpler to setup and captures many nonidealities inherent to the circuit implementation (such as component tolerances), the spectral admittance method allows for greater design flexibility and design space exploration. For this method, the behavior of $C_{ri}'$, $C_m$, and the MEMS switch is abstracted as a time-varying capacitance with a total shunt admittance of $jC_0 Y_{SAM}$ where $Y_{SAM}$ is given by ($\ref{spectral_admittance}$) and $C_0$ is the time-averaged shunt capacitance from $(\ref{center_freq_eqn})$. Similar to \cite{wu_isolating_2019, zhao_design_2023}, the diagonal components in ($\ref{spectral_admittance}$) represent the capacitor's admittance at all fundamental and intermodulation (IM) frequencies $\omega \pm n\omega_m$. The off-diagonal components represent the transadmittance between IM products. $a_k$ represent the Fourier series coefficients of the modulation waveform and, for a square wave, are given by
\begin{equation}
    a_k = \frac{\sin\left(\pi kD\right)}{k\pi}e^{-j2\pi\theta k},
\end{equation}
    
    
    \begin{figure*}[!t]
        \centering
        \begin{equation}
        \label{abcd_cascade_matrix}
        \begin{split}
            \begin{bmatrix}[rr]
                A&B\\C&D
            \end{bmatrix}_{Filter} = &\begin{bmatrix}[rr]
                I & -jC_{k1, s}^{-1}\left(IY_{SAM}\right)^{-1} \\  Z & I
            \end{bmatrix}\begin{bmatrix}[rr]
                I & Z \\ jC_{k1, p} IY_{SAM} & I
            \end{bmatrix}\begin{bmatrix}[rr]
                I & Z \\ jC_0 Y_{SAM} -jL_r^{-1}\left(IY_{SAM}\right)^{-1} & I
            \end{bmatrix}_{\phi_1}\\
            &\begin{bmatrix}[rr]
                I & Z \\ -jC_{k2} IY_{SAM} & I
            \end{bmatrix}
            \cdots\begin{bmatrix}[rr]
                I & Z \\ jC_0 Y_{SAM} -jL_r^{-1}\left(IY_{SAM}\right)^{-1} & I
            \end{bmatrix}_{\phi_i} \cdots
            \begin{bmatrix}[rr]
                I & -jC_{kN, s}^{-1}\left(IY_{SAM}\right)^{-1} \\  Z & I
            \end{bmatrix}
        \end{split}
        \end{equation}
        \vspace{-0.1in}
    \end{figure*}

    \begin{figure*}[b]
        \centering
        \subfloat[]{\includegraphics[height=0.38\textwidth]{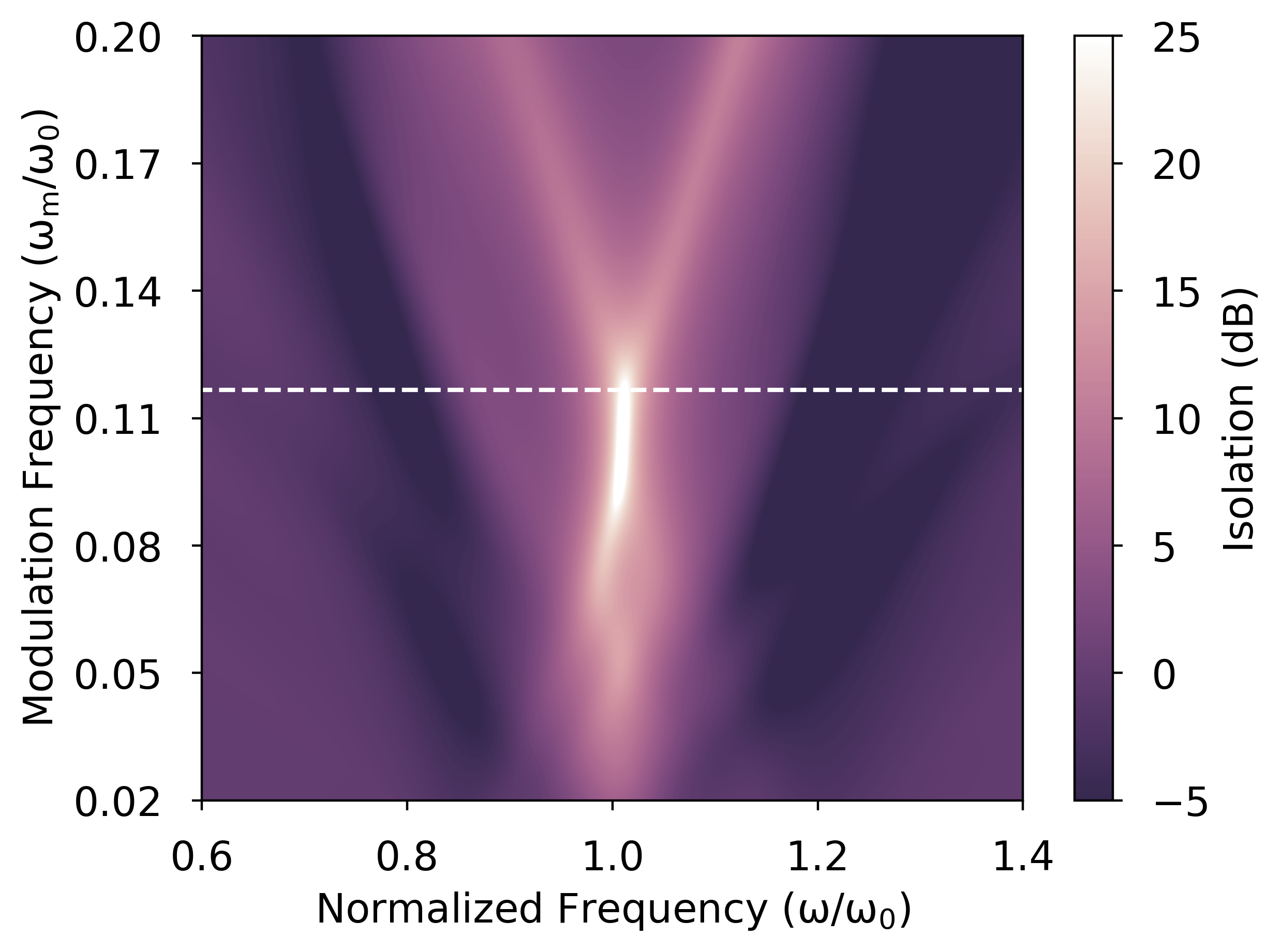}%
        \label{3rdOrder_Ideal_Isolator_ModFreqSweep_BiasLine}}
        \subfloat[]{\includegraphics[height=0.38\textwidth]{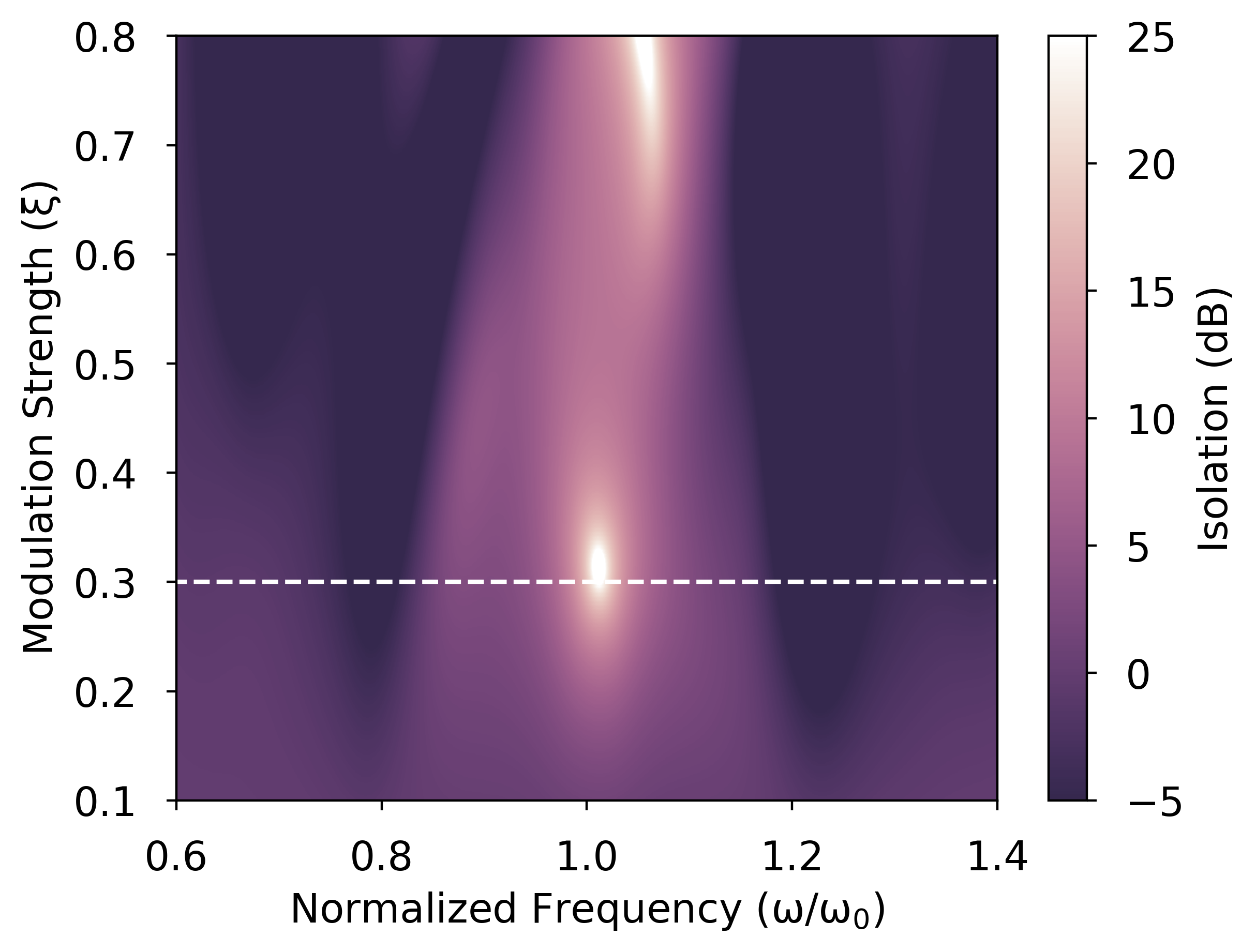}%
        \label{3rdOrder_Ideal_Isolator_ModStrSweep}}
        \caption{Simulated isolation for a $3^{rd}$-order lossless STM isolator as the \textbf{(a)} modulation frequency ($\omega_m$) and \textbf{(b)} modulation strength ($\xi$) are swept with $\phi_m=60^\circ$. In \textbf{(a)} $\xi$ is fixed at $0.3$ and in \textbf{(b)} $\omega_m/\omega_0$ is fixed at $0.1167$. A normalized modulation frequency of $\omega_m/\omega_0=0.1167$ is chosen to maximize isolation and isolation bandwidth while minimizing forward transmission loss.}
        \label{simulated_3rd_order_sweeps}
    \end{figure*}

where $\theta$ is a global phase. Once the SAM is formed, the filter circuit can be simulated using standard network analysis by cascading block $ABCD$-matrices as shown in ($\ref{abcd_cascade_matrix}$) and then converting to a block S-parameter matrix to extract the transmission response. In ($\ref{abcd_cascade_matrix}$), $Z$ is the zero matrix, $I$ is the identity matrix, $\phi_i$ is the modulated phase of the $i^{th}$ resonator in the SAM. Non-modulated frequency-dependent components are represented as block diagonal $ABCD$-matrices using only the diagonal components of ($\ref{spectral_admittance}$) as $IY_{SAM}$. The spectral $ABCD$-matrices are infinite in dimension and must be truncated to $p \times p$ for simulation. The IM-products closest to the fundamental frequency will have the strongest influence on the STM isolator performance so the number of IM-products ($p$) included in the simulation is increased until the simulated isolator response shows minimal change. The data in this paper is simulated using $p=11$ IM-products. The isolator s-parameter matrix at the fundamental frequency is a sub-matrix of final block S-parameter matrix formed by taking the elements at index $(\frac{p+1}{2}, \frac{p+1}{2})$ from each block over frequency.

Through numerical optimization over the space of modulation parameters for an ideal isolating bandpass filter (no resistive loss), it is observed that the strongest isolation occurs when the modulation phase is $\phi_m=\frac{180^\circ}{N}$ where $N$ is the filter order and the modulation duty cycle is $50\%$. The optimal choice of the modulation strength $\xi$ and frequency $\omega_m$ depends strongly on the filter order, ripple parameter, and fractional bandwidth ($\Delta$). Fig. $\ref{simulated_3rd_order_sweeps}$ shows the filter isolation as $\xi$ and $\omega_m$ are swept for a $3^{rd}$-order $\SI{0.05}{\decibel}$ Chebyshev filter with $15\%$ fractional bandwidth and a center frequency of $\omega_0$. In this case, the isolation's sign indicates transmission direction, where a negative value corresponds to strong reverse transmission. In Fig. $\ref{3rdOrder_Ideal_Isolator_ModFreqSweep_BiasLine}$, a lower modulation frequency increases forward transmission loss while a higher modulation frequency reduces the non-reciprocity. To maintain low transmission loss and strong isolation, a normalized modulation frequency of $11.67\%$ is selected. In Fig. $\ref{3rdOrder_Ideal_Isolator_ModStrSweep}$, a higher $\xi$ increases non-reciprocity but also increases transmission loss, so a modulation strength of $\xi=0.3$ is used. It should be emphasized that in these simulations of an ideal isolator, the forward transmission loss is not due to resistive dissipation. Instead, the power is converted into nearby IM-products and is not recovered at the fundamental frequency.

\section{Experimental Results}
    \begin{table} [t]
        \caption{$3^{rd}$-Order $\SI{0.05}{\decibel}$ Chebyshev Isolator Components Values}
        \centering
        \renewcommand{\arraystretch}{1.5}
        \begin{tabular}{|p{1.2cm}|p{3.8cm}|}\hline
            $L_{r}$ ($\times 3$) & $\SI{3.9}{\micro\henry}$\\\hline
            $C'_{r1},C'_{r3}$  & $\SI{10000}{\pico\farad}$\\\hline
            $C'_{r2}$  & $\SI{4300}{\pico\farad}, \SI{5600}{\pico\farad}$\\\hline
            $C_{k1},C_{k4}$  & $\SI{6200}{\pico\farad}$\\\hline
            $C_{k2},C_{k3}$  & $\SI{2700}{\pico\farad}$\\\hline
            $C_{m}$  ($\times 3$) & $\SI{2700}{\pico\farad}$, $\SI{2700}{\pico\farad}$, \textcolor{black}{$\SI{2400}{\pico\farad}$}\\\hline
        \end{tabular}
        \label{component_values}
    \end{table}

    \begin{figure}[b]
        \centering
        \includegraphics[width=0.485\textwidth]{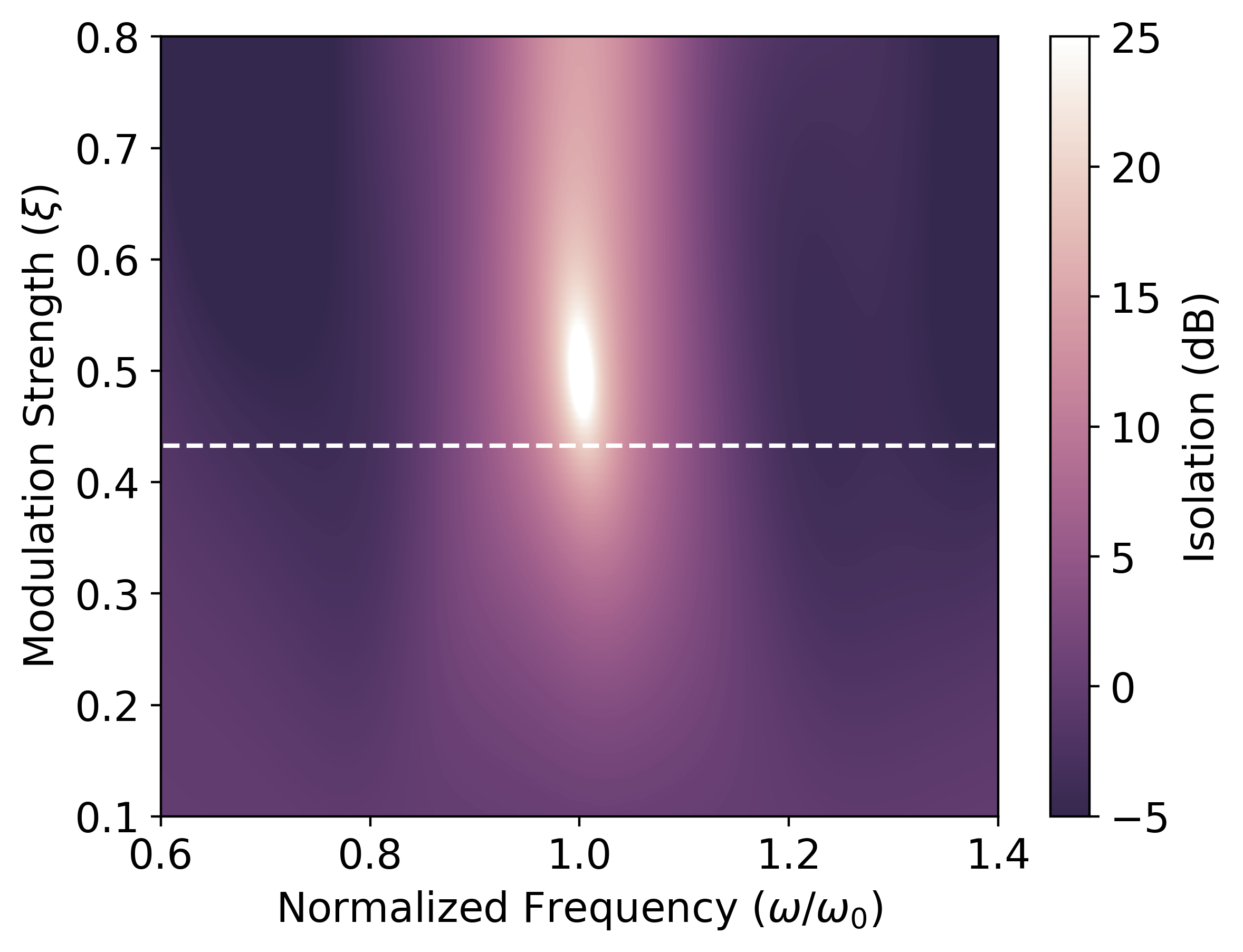}
        \caption{Accounting for component losses in the isolator simulations from the inductors' quality factor ($Q=70$) and resistance ($R_{dc}=\SI{0.824}{\ohm}$) as well as the switches' on-resistance ($R_{switch}=\SI{1.5}{\ohm}$) shows that the required modulation strength for optimal isolation increases by $67\%$ as compared to the lossless case from Fig. $\ref{3rdOrder_Ideal_Isolator_ModStrSweep}$.}
        \label{sim_modstr_sweep}
    \end{figure}
    
    Fig. $\ref{device_photo}$ shows a photo of the $3^{rd}$-order isolator designed using MM5230 MEMS switches from Menlo Microsystems, Inc. The MM5230 switch is rated for stable switching up to $\SI{10}{\kilo\hertz}$. This switching frequency determines the maximum modulation frequency ($\omega_m$) and by extension limits the maximum filter center frequency ($\omega_0$) to $\SI{85}{\kilo\hertz}$ while maintaining high isolation. Using the design techniques from the previous section, the isolator may be realized at any frequency below this $\SI{85}{\kilo\hertz}$ limit by choosing an appropriate switching frequency based on Fig. $\ref{3rdOrder_Ideal_Isolator_ModFreqSweep_BiasLine}$. While this frequency limit may be adequate for many UWAC systems, pursuing a higher maximum center frequency is of interest for a broader range of imaging, sensing, and communication applications. 
    
    To push the center frequency higher, the dynamic response of the Menlo switch is investigated using laser Doppler vibrometry (LDV) while sweeping the frequency of the applied gate waveform. From the LDV results and later confirmed by continuously monitoring the contact resistance as described in \cite{yongbok_cyro_switch} using a resistive divider network, the switch also shows a region of stable operation around $\SI{70}{\kilo\hertz}$. Using $\omega_m=2\pi \cdot \SI{70}{\kilo\hertz}$ enables a maximum center frequency of $\omega_0=2\pi \cdot \SI{0.6}{\mega\hertz}$. Table $\ref{component_values}$ lists the lumped component values used to implement the $3^{rd}$-order isolating bandpass filter from Fig. $\ref{device_photo}$ with a $15\%$ fractional bandwidth at the highest center frequency of $\SI{0.6}{\mega\hertz}$.

    \begin{figure}[b]
        \centering
        \includegraphics[width=0.485\textwidth]{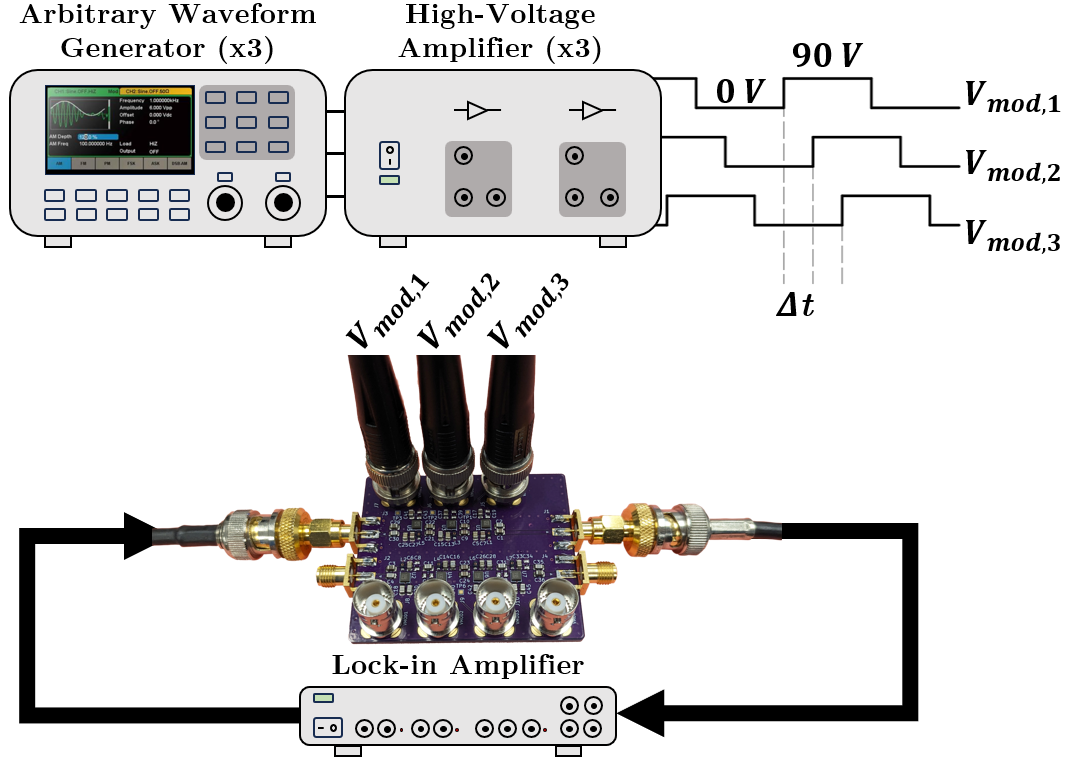}
        \caption{Experimental setup to measure the forward and reverse transmission response of the $3^{rd}$-order STM isolator. A Siglent SDG2042X and a Keysight 33220A arbitrary waveform generators with a shared $\SI{10}{\mega\hertz}$ clock generate the low voltage modulation square waves. Two TEGAM 2350 high-voltage amplifiers are used to increase the modulation voltage above the switches' pull-in voltage to $\SI{90}{\volt}$. A HF2LI Zurich lock-in amplifier measures the STM isolator's transmission response.}
        \label{fig_meas_setup}
    \end{figure}

    The resistive losses and resonator quality factors ($Q$) in the isolator are characterized through measurements of the non-modulated filter with all switches in either the on or off states. An effective switch on-resistance of $R_{switch}=\SI{1.5}{\ohm}$, an inductor Q-factor of $Q=70$, and an inductor DC resistance of $R_{dc}=\SI{0.824}{\ohm}$ together show good agreement with lumped element simulations of the non-modulated filter. Incorporating these loss factors into the spectral admittance simulations shows not only increased forward transmission resistive loss as expected, but also significantly reduced isolation. Re-optimizing the modulation strength in Fig. $\ref{sim_modstr_sweep}$ after accounting for these loss factors illustrates that the isolator must be more strongly modulated in order to compensate for the reduced power of the IM-products from the system's resistive losses. In this case the optimal $\xi$ increases by $67\%$ as compared to the ideal lossless isolator. The PCB design is adjusted to increase $\xi$ from $0.3$ to $0.433$ by adding an extra $\SI{2400}{\pico\farad}$ capacitor to $C_m$.
	 
    \begin{figure*}[t]
        \centering
        \includegraphics[width=0.97\textwidth]{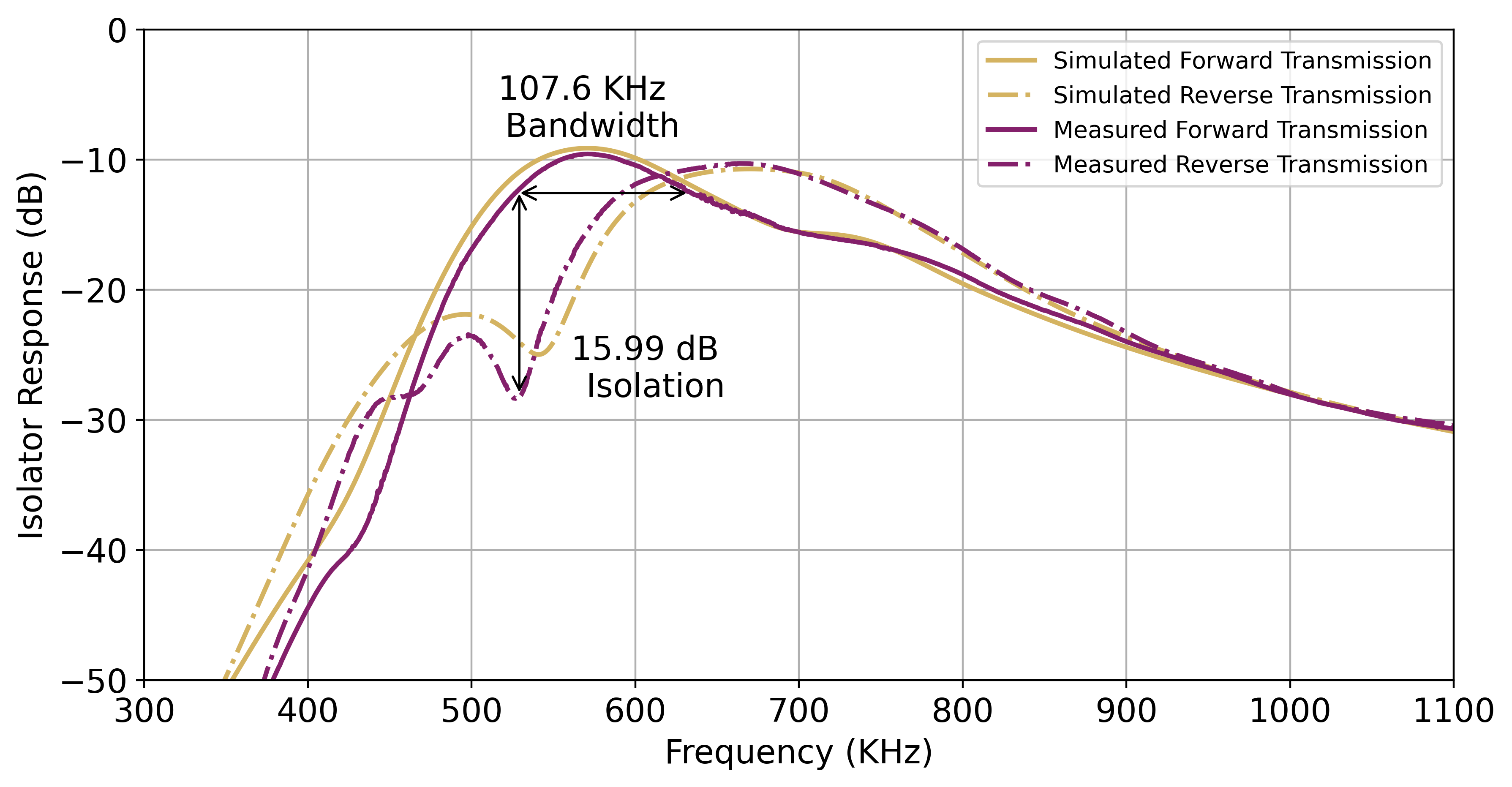}
        \caption{Measured $3^{rd}$-order isolator response compared to simulation. The measurement uses a $\SI{90}{\volt}$ square wave modulation at $\SI{62}{\kilo\hertz}$ with a $60^\circ
        $ phase increment. Simulated response assumes an inductor quality factor of $Q=70$, a resonator series resistance of $R_{dc}=\SI{0.824}{\ohm}$, a switch resistance of $R_{switch}=\SI{1.5}{\ohm}$, and an effective modulation duty cycle of $41\%$. }
        \label{filter_meas_response}
    \end{figure*}

    \begin{figure}[b]
        \centering
        \includegraphics[width=0.485\textwidth]{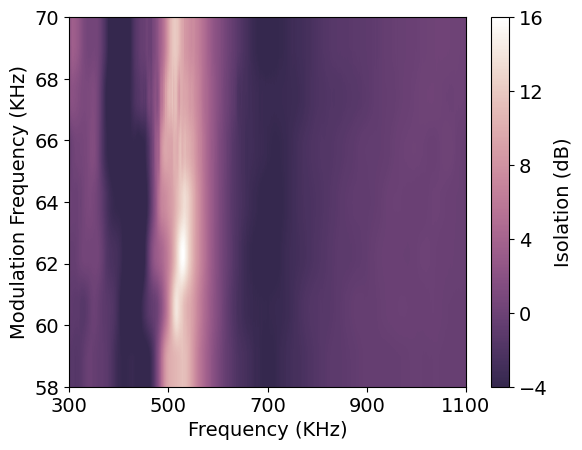}
        \caption{Measured $3^{rd}$-order filter isolation ($\xi=0.43$) over applied modulation frequency. The gate voltage is a $\SI{90}{\volt}$ square wave at $50\%$ duty-cycle modulation with a phase increment of $60^\circ$. The strongest isolation of $\SI{15.99}{\decibel}$ is observed at $\SI{62}{\kilo\hertz}$.}
        \label{filter_meas_response_freq_sweep}
    \end{figure}
    
    The measurement setup used to characterized the modulated filter response is illustrated in Fig. $\ref{fig_meas_setup}$. A Zurich HF2LI lock-in amplifier measures the filter's forward and reverse transmission response over frequency. While the lock-in amplifier is sweeping, three modulation waveforms ($V_{mod,i}$) are applied to the gates of the three MM5230 switches with a relative phase progression of $60^\circ$. A Siglent SDG2042X and
    a Keysight 33220A with a shared $\SI{10}{\mega\hertz}$ clock generate the three low voltage phase locked square waves at $\omega_m$. The low voltage modulation waveforms are amplified beyond the switches' pull-in voltages to $\SI{90}{\volt}$. The forward transmission is measured using $\phi_m=-60^\circ$ while the reverse transmission measurement uses $\phi_m=+60^\circ$. The transmitted voltages measured by the lock-in amplifier are normalized by the frequency response of a short $\SI{50}{\ohm}$ transmission line on a second PCB to isolate the normalized filter response similar to a measurement of $S_{12}$ and $S_{21}$.

\begin{figure}[t]
        \centering
        \includegraphics[width=0.485\textwidth]{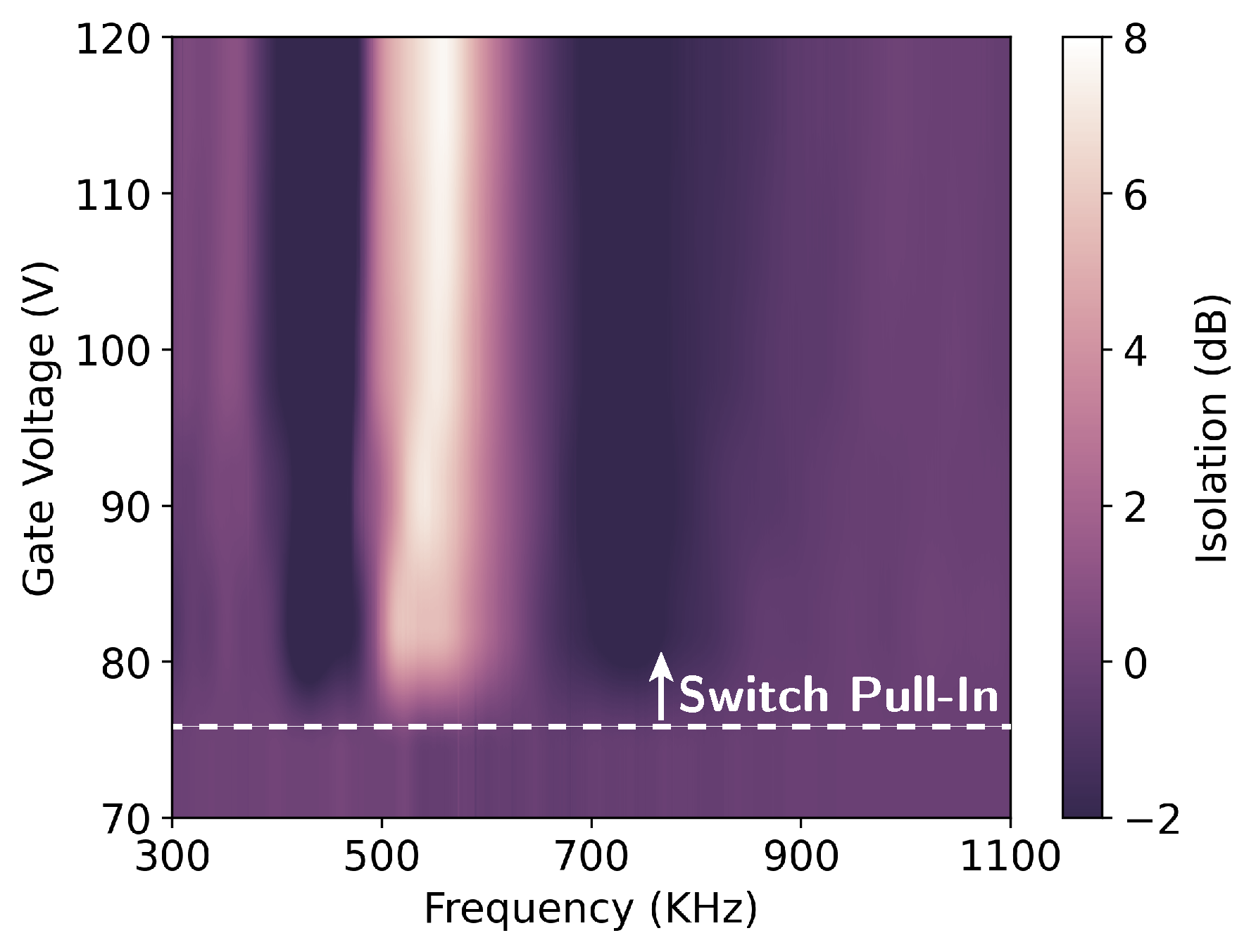}
        \caption{Measured $3^{rd}$-order filter isolation ($\xi=0.30$) over the gate voltage magnitude. The gate voltage a $50\%$ duty-cycle square wave at $\SI{62}{\kilo\hertz}$ with a phase increment of $60^\circ$. Below the switches' pull-in voltage, the filter becomes reciprocal as expected.}
        \label{filter_meas_response_gateV_sweep}
    \end{figure}

    Fig. $\ref{filter_meas_response}$ shows the measured modulated isolator response using $\SI{90}{\volt}$ square wave modulation at $\SI{62}{\kilo\hertz}$, a gate voltage duty cycle of $50\%$, and $\phi_m=\pm 60^\circ$. The forward and reverse transmission responses show excellent agreement with the SAM filter simulation assuming an effective modulation duty cycle of $41\%$ and demonstrate a maximum isolation of $\SI{15.99}{\decibel}$. Fig. $\ref{filter_meas_response_freq_sweep}$ shows the measured filter isolation over frequency while sweeping the modulation waveforms' frequency. The filter demonstrates strong isolation over the entire range with maximum isolation at $\omega_m=2\pi \cdot \SI{62}{\kilo\hertz}$. Since the non-reciprocity is originating from the periodic addition of $C_m$ to the filter's LC-tank resonators, the filter should be non-reciprocal for any modulation waveform magnitude above the switches' pull-in voltages. As illustrated in Fig. $\ref{filter_meas_response_gateV_sweep}$, the filter does show strong non-reciprocity for all modulation voltages above $\SI{80}{\volt}$. At $\SI{80}{\volt}$, the filter isolation is reduced indicating some of the switches within the MM5230 are not closing completely resulting in increased on-resistance (discussed in the appendix). Below $\SI{80}{\volt}$, at least two of the three MM5230 chips do not close and the filter becomes reciprocal as expected. The maximum isolation in Fig. $\ref{filter_meas_response_gateV_sweep}$ is lower than in Fig. $\ref{filter_meas_response_freq_sweep}$ because the measurements used a modulation strength of $\xi=0.3$ instead of $\xi=0.433$.

\section{Discussion}

    \begin{figure}[b]
        \centering
        \includegraphics[width=0.485\textwidth]{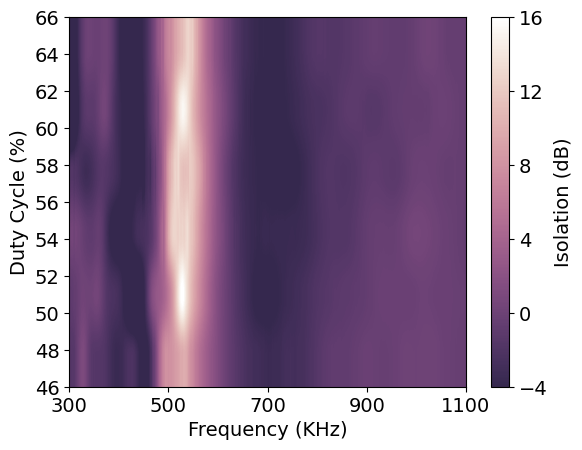}
        \caption{Measured $3^{rd}$-order filter isolation ($\xi=0.433$) over the gate voltage duty cycle. The gate voltage is a $\SI{90}{\volt}$ square wave at $\SI{62}{\kilo\hertz}$ with a phase increment of $60^\circ$. }
        \label{filter_meas_response_DC_sweep}
    \end{figure}
    
Optimal spatiotemporally modulated isolator performance for this PCB implementation is achieved when the effective modulated capacitance duty cycle is $50\%$ and the maximum isolation occurs at the isolator's center frequency. However, from Fig. $\ref{filter_meas_response}$, it is clear that the maximum isolation occurs below the isolator's center frequency. Furthermore, the best fit between measurement and simulation occurs at an effective modulation duty cycle of $41\%$ despite an input gate voltage duty cycle of $50\%$. Sweeping the duty cycle of the applied modulation waveforms in Fig. $\ref{filter_meas_response_DC_sweep}$ shows minimal improvement in the filter's isolation.

    \begin{figure}[t]
        \centering
        \subfloat[]{\includegraphics[width=0.485\textwidth]{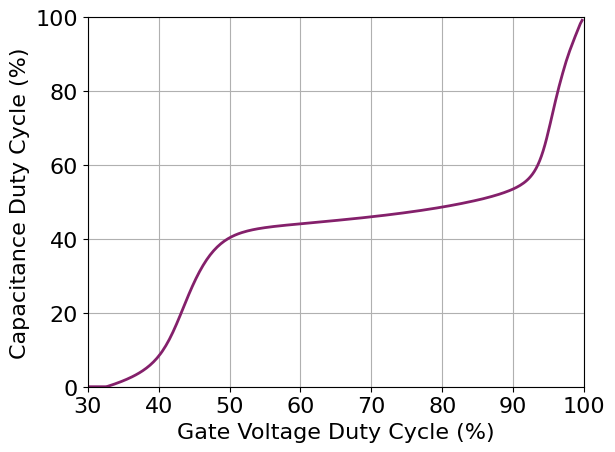}%
        \label{CapacitanceDutyCycle_Sim}}
        
        \subfloat[]{\includegraphics[width=0.485\textwidth]{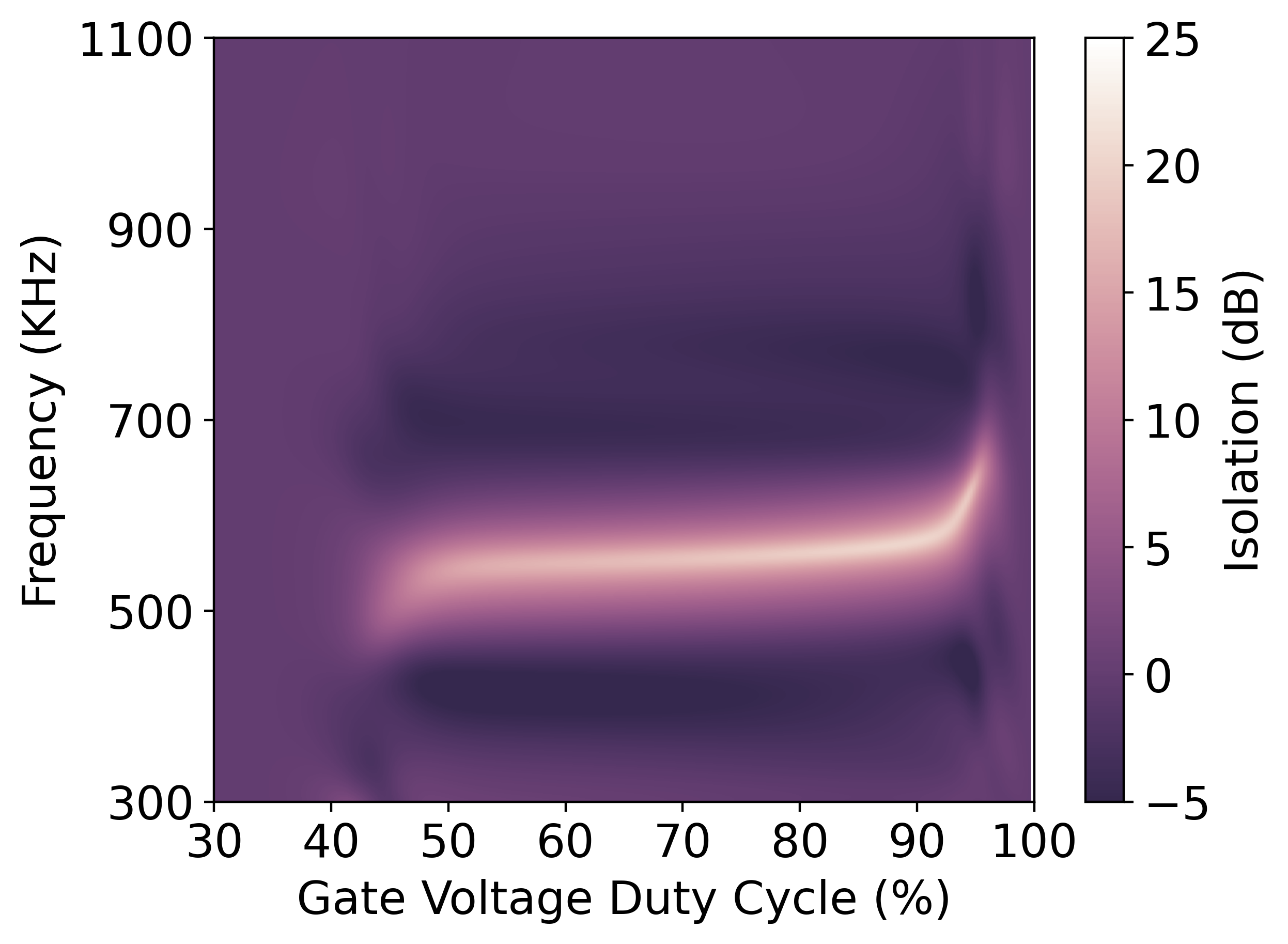}%
        \label{CapacitanceDutyCycle_Sim_Isolation}}
        \caption{\textbf{(a)} The beam dynamics of the Menlo switch are simulated numerically to analyze the effective capacitance duty cycle of $C_m$ over the applied gate voltage square wave duty cycle. \textbf{(b)} Resulting isolation over the input duty cycle.}
        \label{beam_dynamics}
    \end{figure}
    
To investigate the source of this discrepancy, the mechanical switch is modeled as a single cantilever beam using a standard 1-dimensional spin-mass-damper system, incorporating a capacitive electrostatic forcing function \cite{kaajakari2009practical}:
\begin{equation}
    \frac{dv(t)}{dt} = \frac{1}{m}\left[ \frac{1}{2}\frac{V_g(t)^2\varepsilon A}{(g-x(t))^2 } + f_s(x(t)) - \zeta v(t) - kx(t)\right],
\end{equation}
where $v(t)$ and $x(t)$ are the beam velocity and position respectively. $m$ is the beam's effective mass, $k$ is the spring constant, $\zeta$ is the damping parameter, $A$ is the effective area of the gate electrode, $f_s$ is an additional surface force when the switch is closed \cite{kwon_investigation_2008}, $V_g(t)$ is the gate voltage, and $g$ is the nominal distance between the beam and switch contact. 
The dynamic switch response is simulated using an adaptive-step Runge-Kutta solver which pauses integration on each collision with the switch contact to account for the cantilever bouncing \cite{peschot_contact_2012}. The simulations use a mechanical resonant frequency of $\SI{133}{\kilo\hertz}$, a Q-factor of $5.13$, a pull-in voltage of $\SI{70.4}{\volt}$, a pull-off voltage of $\SI{63.0}{\volt}$, and a coefficient of restitution of $-0.22$ which are obtained through LDV measurement. The total time-varying resonator capacitance ($C_{ri}' + C_m$) is simulated over a range of input gate voltage duty-cycles with $\omega_m=2\pi\cdot \SI{70}{\kilo\hertz}$. The resulting modulation duty cycle is extracted from the capacitance time series data and plotted in Fig. $\ref{CapacitanceDutyCycle_Sim}$. This effective modulation capacitance is then used in the SAM isolator simulations to illustrate the isolation over input gate voltage duty cycle in Fig. $\ref{CapacitanceDutyCycle_Sim_Isolation}$. From these results, the effective modulation duty cycle is insensitive to the gate voltage duty cycle around $50\%$ aligning well with the measured results in Fig. $\ref{filter_meas_response_DC_sweep}$. Above $85\%$ duty cycle, the maximum isolation improves and the frequency of maximum isolation shifts higher to the center of the passband as anticipated.

Practically, driving the switches at such a high duty cycle is challenging since the effective modulation is very sensitive to the input duty cycle and the shape of Fig. $\ref{CapacitanceDutyCycle_Sim}$ is sensitive to the switch's mechanical properties such as $Q$-factor, resonant frequency, and pull-in voltage which vary from device to device. Instead, if the spatiotemporal isolator and modulation parameters are designed to provide maximum isolation between a gate voltage duty cycle of $40\%$ and $50\%$, then the isolator performance will be largely insensitive to device fabrication variations maintaining near optimal isolation. Additionally, reducing the filter's resistive losses from the chip inductors and the switch contact will provide significant enhancement in the forward transmission loss and isolation simultaneously.  
    

\section{Conclusion}

    \begin{table} [b]
        \caption{Comparison with the State-of-the-Art}
        \centering
        \renewcommand{\arraystretch}{1.5}
        \begin{tabular}{|p{0.6cm}|p{1.2cm}|p{1.2cm}|p{1.2cm}|p{1.0cm}|p{1.0cm}|}\hline
             & Center Frequency & Bandwidth ($\%$) & Insertion Loss & Isolation & $\omega_m/\omega_0$ \\\hline
            This Work  & $\SI{570.8}{KHz}$ & $18.85\%$ & $\SI{9.56}{\decibel}$ & $\SI{15.99}{\decibel}$ & $11.7\%$\\\hline
            \cite{lu_radio_2019} & $\SI{155}{\mega\hertz}$ & $8.8\%$ & $\SI{6.6}{\decibel}$ & $\SI{25.4}{\decibel}$ & $0.56\%$ \\\hline
            \cite{wu_isolating_2019} & $\SI{200}{\mega\hertz}$ & $14\%$ & $\SI{1.5}{\decibel}$ & $\SI{20}{\decibel}$ & $14\%$ \\\hline
            \cite{yu_magnetic-free_2018} & $\SI{900}{\mega\hertz}$ & $2.6\%$ & $\SI{2.9}{\decibel}$ & $\SI{20}{\decibel}$ & $1.4\%$ \\\hline
            \cite{wu_non-reciprocal_2019} & $\SI{1.0}{\giga\hertz}$ & $6.3\%$ & $\SI{5.5}{\decibel}$ & $\SI{6.2}{\decibel}$ & $4.6\%$ \\\hline
            \cite{pirro_low_2019} & $\SI{1175}{\mega\hertz}$ & $1.7\%$ & $\SI{4.5}{\decibel}$ & $\SI{28}{\decibel}$ & $3.4\%$ \\\hline
            \cite{khater_switch-based_2024} & $\SI{1.3}{\giga\hertz}$ & $4.3\%$ & $\SI{5.9}{\decibel}$ & $\SI{20}{\decibel}$ & $1.5\%$\\\hline
        \end{tabular}
        \label{sota_table}
    \end{table}

This work presents the design and implementation of a magnet-free, low-frequency spatiotemporally modulated isolator using MEMS switches for self-interference cancellation in underwater acoustic communication systems. By leveraging the high linearity and power handling characteristics of Menlo Microsystems switches, the proposed design enables in-band full duplex operation (when combined with digital self-interference cancellation techniques) for frequencies below $\SI{0.6}{\mega\hertz}$ where conventional circulators are unavailable. The printed circuit board (PCB) implementation shows a maximum isolation of $\SI{15.99}{\decibel}$ and excellent agreement with the results from the spectral admittance matrix simulation method. Furthermore, there is a clear path toward improving isolation and achieving isolator performance comparable to the state-of-the-art at higher frequencies (Table $\ref{sota_table}$) by optimizing the effective modulation duty cycle, minimizing resistive losses, and increasing the isolator order (as discussed in the appendix). The design curves presented in this study outline a straightforward procedure for realizing isolators at any frequency below $\SI{0.6}{\mega\hertz}$, targeting a wide range of UWAC applications.

One potential concern for a MEMS switch based STM isolator implementation is the gradual creep of the mechanical cantilever beam eventually resulting permanent striction and a loss of non-reciprocity. Based on the recent studies of this failure mechanism \cite{gu_lifetime_2024}, cryogenic testing \cite{yongbok_cyro_switch}, and the lifetime testing in the MM5230 datasheet \cite{micro_rf_nodate}, the switch will reliably operate for at least 3 billion cycles. During the data collection in this work, no creep lifetime related switch failures were observed.

Unlike the conventional ferrite circulator, this STM isolator offers a unique capability to dynamically change its non-reciprocity through control of the modulation waveforms.  For example, the isolation can be disabled by turning off the modulation, the transmission direction can be reversed by flipping the modulation phase progression, and, based on Fig. $\ref{CapacitanceDutyCycle_Sim_Isolation}$, the frequency of maximum isolation can be fine tuned by adjusting the gate voltage duty cycle. These features of the STM isolator have the potential to enable novel adaptive communication system architectures.

\section*{Appendix}
    \begin{figure}[b]
        \centering
        \includegraphics[width=0.485\textwidth]{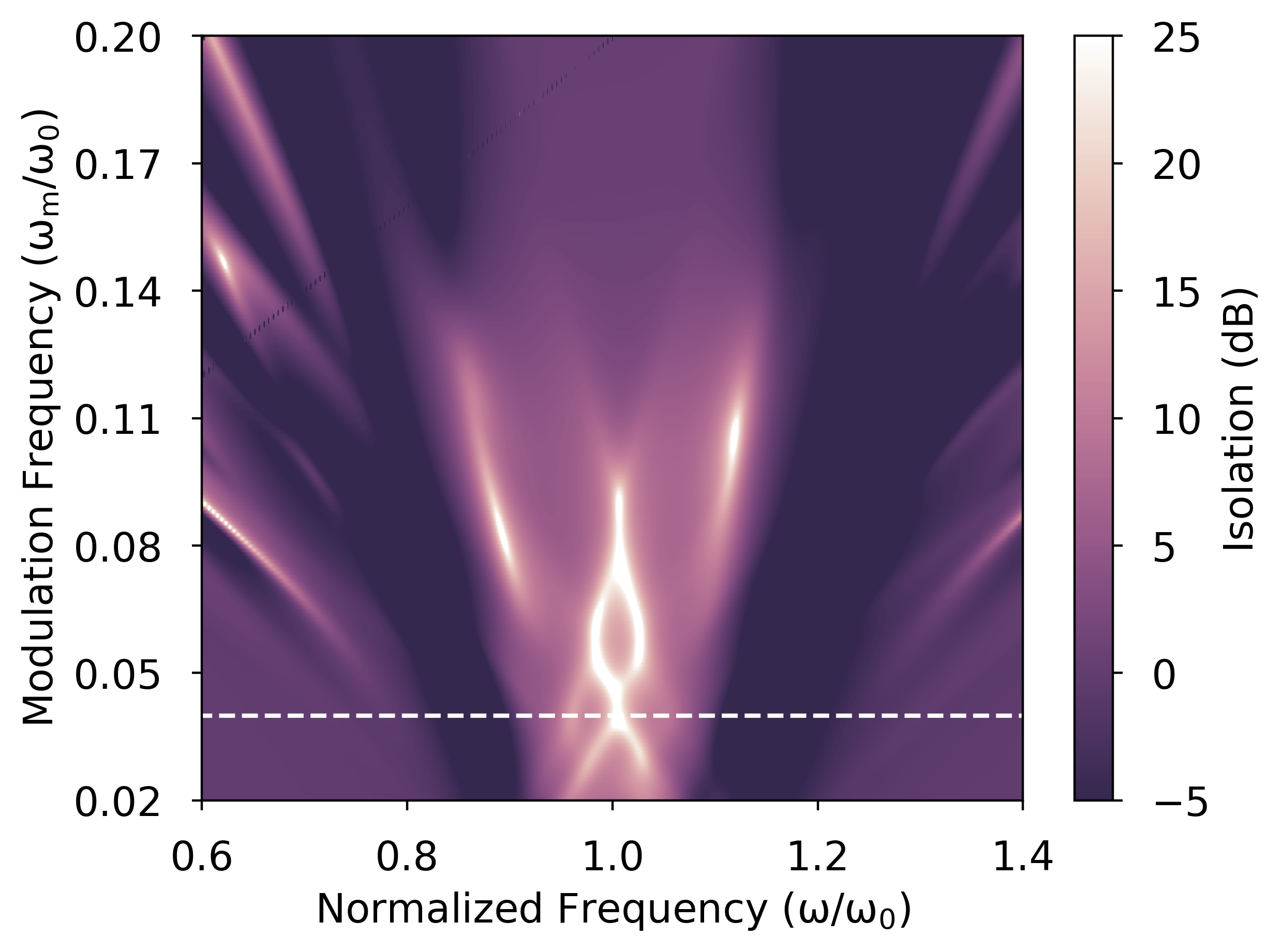}
        \caption{Simulated isolation for a $5^{th}$-order $\SI{0.01}{\decibel}$ Chebyshev lossless STM isolator as the modulation frequency ($\omega_m$) is swept. $\Delta=0.15$, $\xi=0.16$, $\phi_m=36^\circ$, and the effective modulation duty cycle is $50\%$. At $\omega_m/\omega_0=0.04$, the maximum isolation is $\SI{39.2}{\decibel}$.}
        \label{5thOrder_Ideal_Isolator_ModFreqSweep_BiasLine}
    \end{figure}
    
The lifetime and power consumption of the STM isolator will be key figures of merit for its adoption into many applications. The power dissipated to charge and discharge the gate electrode while actuating the MEMS switch is $P\propto C_{gate} V^2 f_m$ and the switch's lifetime is inversely proportional to the modulation frequency $f_m$. To simultaneously improve both of these figures, it is advantageous to reduce the required modulation frequency while maintaining the same center frequency and isolation. Increasing the filter order provides one method to reduce $\omega_m/\omega_0$ while preserving high isolation. A larger number of modulated resonators provide additional degrees of freedom to control the mixing of IM-products, and, for an optimized set of modulation parameters, can provide a significant enhancement. For example, if the filter order is increased from $3^{rd}$-order to $5^{th}$-order, then $\omega_m/\omega_0$ can be reduced from $0.1167$ to $0.04$ while maintaining over $\SI{39}{\decibel}$ of isolation as shown in Fig. $\ref{5thOrder_Ideal_Isolator_ModFreqSweep_BiasLine}$ for a lossless STM isolator.

    \begin{figure}[t]
        \centering
            \begin{circuitikz}[scale=0.6]
                \ctikzset{label/align=rotate}
                
                \ctikzset{capacitors/scale=0.4}
                \ctikzset{resistors/scale=0.5}
                \ctikzset{inductors/scale=0.5}
                \ctikzset{bipole label style/.style={font=\footnotesize}}
                
            
                \draw (3,-6) to (3,-6) node[ground]{}; 
                
                \draw(7,-1) -- (8,-1) node[fill=white, scale=2]{\(\cdots\)};
                
                \draw (4,-3.5) to[closing switch, switch start arrow=latexslim,
                l_=$\theta_1$] (4,-2.5);
                \draw (6,-3.5) to[closing switch, switch start arrow=latexslim,
                l_=$\theta_2$] (6,-2.5);
                \draw (8,-3.5) to[closing switch, switch start arrow=latexslim,
                l_=$\theta_3$] (8,-2.5);
                
                \draw
                
                (0,-1) to[C=$C_{k1}$] (3,-1)
                (3,-1) to[C=$C_{k2}$] (7,-1)
                
                (1.75,-6) to[L=$L_{r}$] (1.75,-2)
                (3,-6) to[C=$C_{r1}'$] (3,-2)
                (4,-2) to[short, -o] (4,-2.5)
                (4,-4) to[short, -o] (4,-3.5)

                (6,-2) to[short, -o] (6,-2.5)
                (6,-4) to[short, -o] (6,-3.5)

                (8,-2) to[short, -o] (8,-2.5)
                (8,-4) to[short, -o] (8,-3.5)
                
                (4,-6) to[C=$C_{m}$] (4,-4)
                (6,-6) to[C=$C_{m}$] (6,-4)
                (8,-6) to[C=$C_{m}$] (8,-4)

                (1.75,-2) -- (8,-2)
                (1.75,-6) -- (8,-6)
                (3,-1) -- (3,-2)

                ;
            \end{circuitikz}
        \caption{Circuit diagram for a subsection of the lumped-element $3^{rd}$-order STM isolator from Fig. $\ref{filter_schematic}$ modified to use harmonic beamforming. Here the MEMS switch is controlling three distinct modulation capacitors. The gate voltage for each branch has a phase $\theta_i$ relative to $\theta_1$.}
        \label{HB_filter_schematic}
    \end{figure}
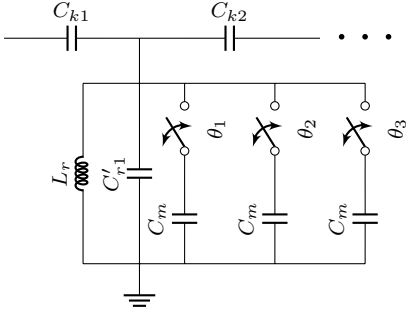

    \begin{figure}[b]
        \centering
        \includegraphics[width=0.485\textwidth]{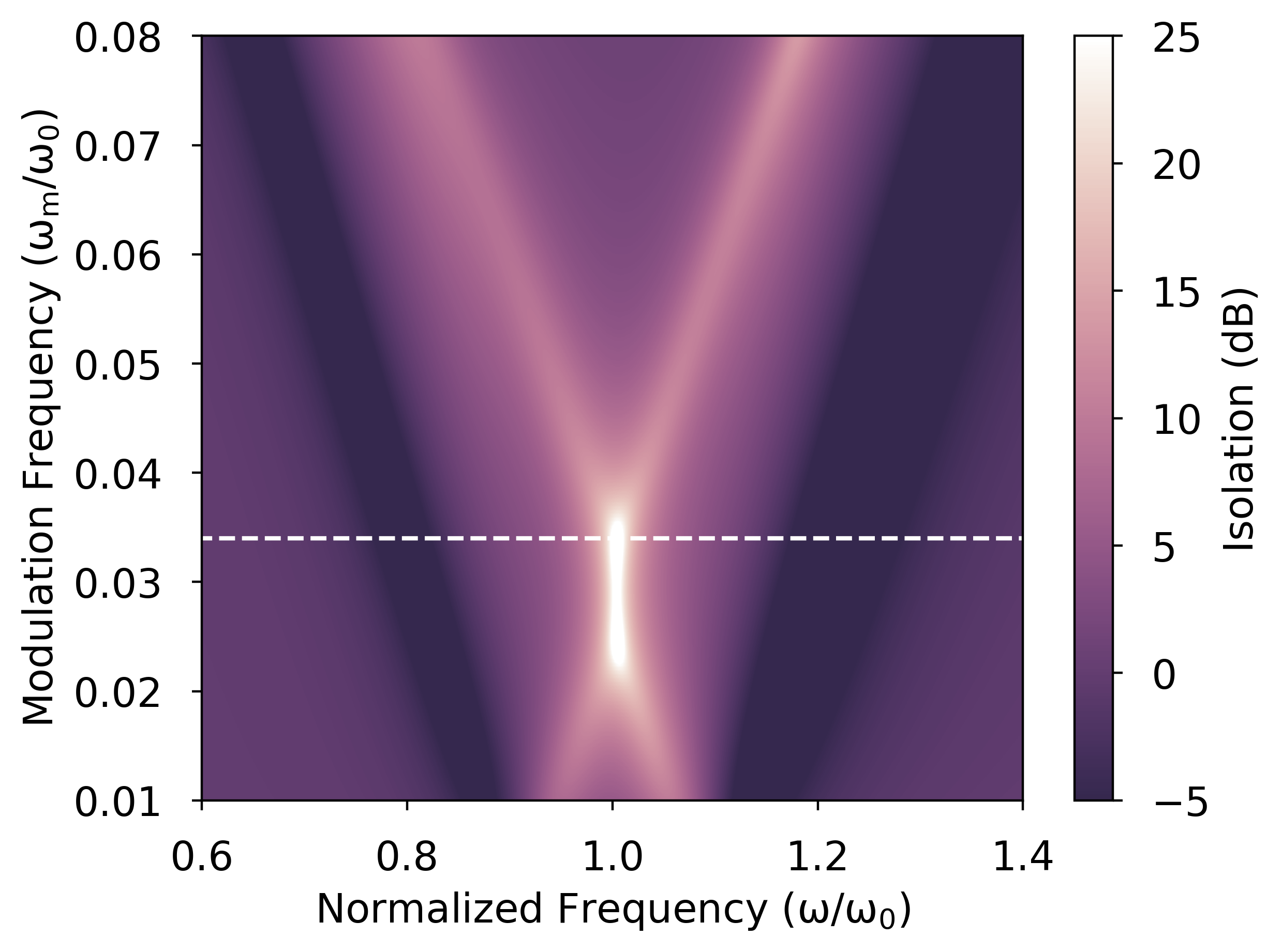}
        \caption{Simulated isolation for a $3^{rd}$-order $\SI{0.05}{\decibel}$ Chebyshev lossless STM isolator with harmonic beamforming using three modulated capacitors as the modulation frequency ($\omega_m$) is swept. $\Delta=0.15$, $\xi=0.30$, $\phi_m=20^\circ$, and the effective modulation duty cycle is $50\%$. $\theta_2=\frac{1}{3}$ and $\theta_3=\frac{2}{3}$ relative to $\theta_1$ giving an effective modulation frequency of $3\omega_m$. At $\omega_m/\omega_0=0.034$, the maximum isolation is $\SI{47.6}{\decibel}$.}
        \label{HBN_3rdOrder_Ideal_Isolator_ModFreqSweep_BiasLine}
    \end{figure}
    
Once resistive losses in the switches and in the resonators are taken into consideration, increasing the isolator order will provide diminishing returns. The single-pole 4-throw nature of the Menlo Microsystems MEMS switches provides a unique opportunity to perform harmonic beamforming, giving a similar reduction in $\omega_m$ without introducing additional loss. In Fig. $\ref{filter_schematic}$, each MM5230 chip is configured in a super-port configuration where a single gate voltage waveform shorts 2 ports of the switch together and adds $C_m$ to the filter circuit. The remaining 2 ports on the chip remain unused. Instead, each of the throws in series with a modulation capacitor can be individually controlled as depicted in Fig. $\ref{HB_filter_schematic}$. In this case, the Fourier series coefficients used in the SAM from ($\ref{spectral_admittance}$) are given by
    \begin{equation}
        \label{HB_array_factor}
        a_k = \frac{\sin\left(\pi kD\right)}{k\pi} \left(1 + e^{-j2\pi\theta_2 k} +  e^{-j2\pi\theta_3 k}\right),
    \end{equation}
where $\theta_2$ and $\theta_3$ are the gate voltage phases relative to $\theta_1$. In an analogy to antenna array processing, equation $(\ref{HB_array_factor})$ can be viewed as the product of an element factor and an array factor \cite{balanis_antenna_2005}. Through proper choice of gate voltage phases $\theta_i$ and the duty cycle $D$, certain transadmittance terms in the SAM can be strongly suppressed while directing energy towards IM-products of interest. For example, with $\theta_1=0$, $\theta_2=\frac{1}{3}$, and $\theta_3=\frac{2}{3}$, then transadmittance terms at $a_1$ and $a_5$ as well as all even ordered terms are suppressed and the energy is directed towards $a_3$ giving an effective modulation at $3\omega_m$. Fig. $\ref{HBN_3rdOrder_Ideal_Isolator_ModFreqSweep_BiasLine}$ illustrates the isolation for a $3^{rd}$-order STM circuit using this harmonic beamforming method while $\omega_m$ is swept. By comparison with Fig. $\ref{3rdOrder_Ideal_Isolator_ModFreqSweep_BiasLine}$, the required modulation frequency is reduced by a factor of 3 and the isolator with harmonic beamforming shows an isolation over $\SI{47}{\decibel}$ at  $\omega_m/\omega_0=0.034$. Notably, the modulation phase increment must be reduced by a factor of $3$ to $\phi_m=20^\circ$ as compared with the case without harmonic beamforming ($\phi_m=60^\circ$). Using harmonic beamforming in conjunction with increased isolator order can simultaneously increase the isolator lifetime by an order of magnitude and reduce the power consumption to drive the gate voltages.\\

    \begin{figure}[b]
        \centering
        \includegraphics[width=0.485\textwidth]{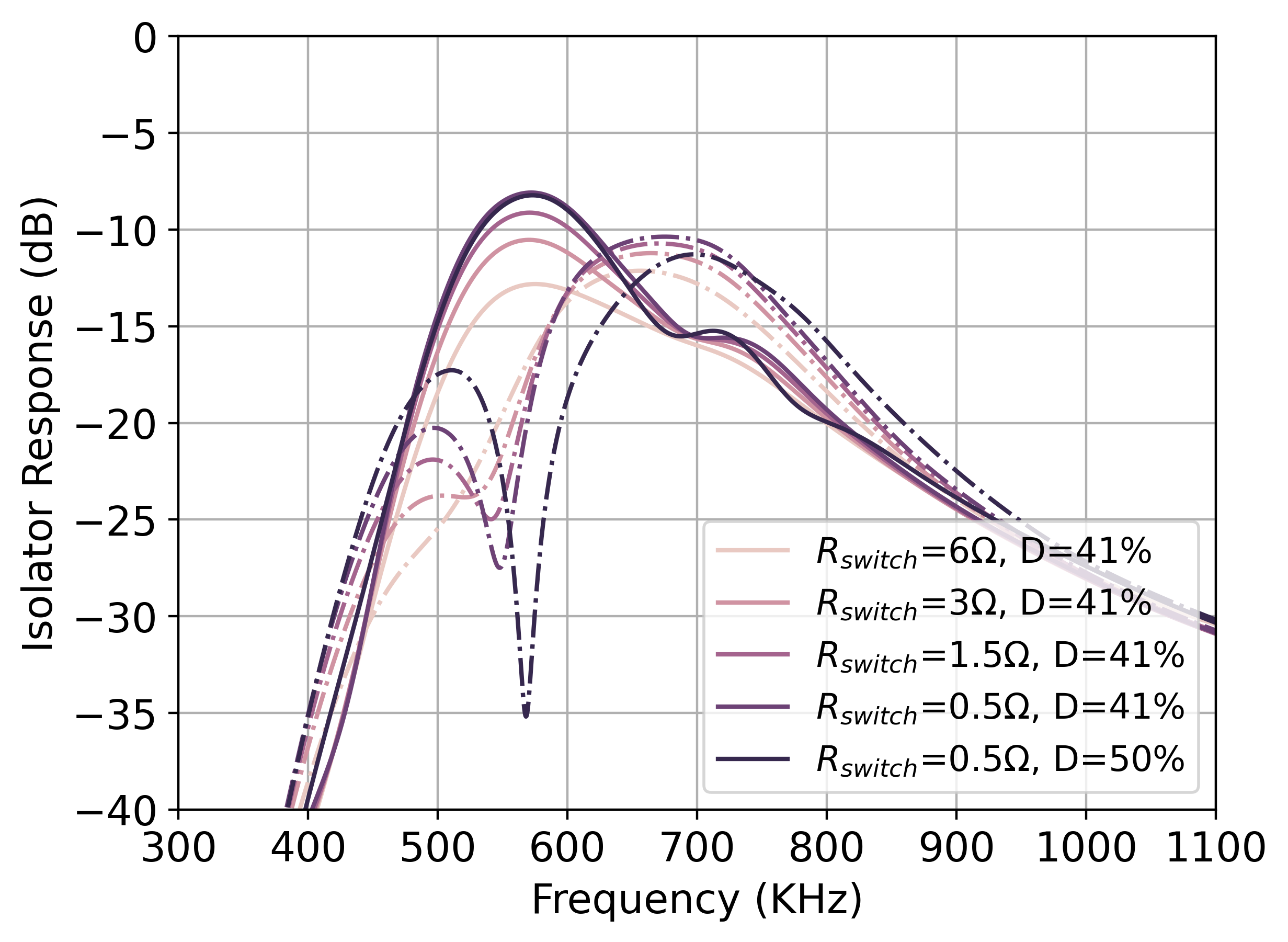}
        \caption{Simulated forward and reverse (dashed) transmission for a lossy STM isolator using an inductor $Q=70$ and resistance $R_{dc}=\SI{0.824}{\ohm}$ as the switch's on-resistance ($R_{switch}$) is sweep from $\SI{6}{\ohm}$ down to $\SI{0.5}{\ohm}$.}
        \label{3rdOrder_Lossy_62KHz_SwitchResistance}
    \end{figure}
    
The losses from the lumped element PCB implementation such as inductor's resistance and Q-factor degrade both forward transmission loss and isolation as discussed in Fig. $\ref{sim_modstr_sweep}$. The MEMS switch's on-resistance also has a strong influence on the isolation as illustrated by the simulation in Fig. $\ref{3rdOrder_Lossy_62KHz_SwitchResistance}$. As the switch's resistance reduces from $\SI{6}{\ohm}$ to $\SI{0.5}{\ohm}$ (using an effective duty cycle of $D=41\%$), the maximum isolation improves from $\SI{7.9}{\decibel}$ to $\SI{18.9}{\decibel}$. Reducing $R_{switch}$ beyond $\SI{0.5}{\ohm}$ provides at most only an additional $\SI{3}{\decibel}$ of isolation. Instead, the isolation can be dramatically improved as illustrated in Fig. $\ref{beam_dynamics}$ by increasing the effective modulation duty cycle close to $50\%$. In this case with $R_{switch}=\SI{0.5}{\ohm}$, the isolation increases to $\SI{26.9}{\decibel}$.

\section*{Data Availability}
        The isolator measurement data produced during this study will be released in a Zenodo repository upon publication.

\section*{Acknowledgments}
    PCB assembly and isolator measurements were performed at Seng-Liang Wang Hall at Purdue. Switch characterization was performed at the Birck Nanotechnology Center, the mechanical engineering building at Purdue, and at Menlo Microsystems, Inc. in Albany, NY. 
    
    This research was developed with funding provided by Menlo Microsystems, Inc.


    \balance
    
\bibliographystyle{IEEEtran}
\bibliography{isolator_bib}
    
	\vfill
    \balance

\end{document}